\documentstyle[preprint,aps,eqsecnum]{revtex}
\begin{document}
\title {Non Fermi liquid renormalization of the conductivity
of fermions coupled to gauge fields}

\author {D.V.Khveshchenko$^{1}$ and Michael Reizer$^{2}$}

\address {$^{1}$ NORDITA, Blegdamsvej 17, Copenhagen DK-2100, Denmark\\
$^{2}$ Department of Physics, The Ohio State University, Columbus, \\
OH 43210-1106}

\maketitle

\begin{abstract}
\noindent

The method of the quantum kinetic equation is applied to the problem of renormalization 
of the conductivity of 
normal metals by gauge electron-electron interactions. It is shown
that in the three-dimensional case the relativistic electromagnetic  
interaction (vector interaction of electrons with transverse photons) 
leads to an unusual temperature dependence, 
indicating a deviation from the Fermi liquid theory at 
low temperatures. In two dimensions such corrections are found to
result from both the scalar (density-density or Coulomb) and the vector
(current-current) gauge interactions.

\end{abstract}
\pagebreak

\section{Introduction}
The renormalization of different electronic properties due to the 
electron-electron interactions constitutes one of the major premises of the Fermi 
liquid theory (FLT). 
Although there exists lots of results about renormalizations of such
equilibrium thermal quantities as specific heat or
Pauli magnetic susceptibility \cite{L,PN}, there are very few statements 
concerning renormalizations of electronic kinetic
coefficients \cite{G,PK}.
This issue was commonly believed to had been settled over 30 years ago
in the work by Prange and Kadanoff \cite{PK}
who considered the conventional electron-phonon problem
and
concluded that neither the electrical conductivity nor the
thermoelectric power are affected by the electron-phonon interactions.

However, it was recently pointed out \cite{SRW} that
the analysis \cite{PK} was not complete. It was stated in \cite{SRW}
that in general
there exists another type of renormalization corrections, which were not taken into account by the authors of \cite{PK}
who only discussed the phonon renormalization of the electron-phonon scattering rate.
The new term derived in \cite{SRW}
can be viewed as resulting from a quantum interference between electron-phonon and electron-impurity scattering
processes. In the framework of the quantum kinetic equation, which we are going to use throughout this paper,
it originates from corrections to the nonequilibrium electron density of states and from the nonlocal
part of the electron-phonon collision integral. 

Unfortunately, the final conclusions made in \cite{SRW} in the context
of the three-dimensional (3D) electron-phonon problem are incorrect.
A more thorough analysis shows that
the new renormalization effect proposed in \cite{SRW} is negligible in all cases of
3D scalar (density-density) interactions, including the 3D electron-phonon, electron-paramagnon, and electron-electron
Coulomb interactions. Nevertheless, as we will show
below, the cancellation which occurs  in the case of the 3D scalar coupling
is not a generic property, but it is only due to 
some pecularity of the relevant phase volume-type expression.  

To this end, in the present paper we start out with an example of 3D vector (current-current) coupling provided by
relativistic electron-electron interactions (the vector interaction of electrons 
with transverse photons), which is known to cause a quite unusual 
and unexpected in FLT behavior of electron quantities at low enough temperatures. Namely, it was previously shown
that the electron energy 
relaxation time $\tau_\epsilon^{-1}$ is proportional to $T\ln T$ \cite{R1},
which obviously violates the FLT criterion of the existence of well-defined fermionic
quasiparticles ($\tau^{-1}_\epsilon <<T$). Moreover, the real part of the 
electron self-energy shows the "marginal"
behavior ${\rm Re}\Sigma(\epsilon)\propto \epsilon \ln \epsilon$ \cite{R2,BMPR}.

The breakdown of the FLT manifests itself in singular corrections
to various thermodynamic quantities,
such as the 
electronic specific heat: $\Delta C_p\propto -T\ln T$ \cite{HNP,R2}.

Generically, the new renormalization
conductivity correction derived in \cite{SRW} can be related to the derivative $\lambda =-{\partial{\rm Re}\Sigma(\epsilon)\over \partial\epsilon}$
(as opposed to the so-called kinetic terms, which can be expressed in terms of 
${\rm Im}\Sigma(\epsilon)$, provided a scattering is quasi-elastic).

In FLT $\lambda (\epsilon\rightarrow 0)$ is proportional to
a coupling constant  and independent of $\epsilon$, which yields a simple
multiplicative reduction of the Drude conductivity  via the effective mass
inhancement: $m=m_0(1+\lambda)$.

If, on the contrary, the function $\lambda (\epsilon)$ becomes singular at
small $\epsilon$, one might expect that the corresponding renormalization
correction to the conductivity
will exhibit such non-Fermi liquid features
as a non-analytic dependence on the coupling strength
and/or temperature.   

In the rest of the paper we demonstrate that in the presence
of both scalar and vector gauge interactions this is indeed the case.

\section{Quantum Kinetic Equation}
We are going to study the renormalization of the classical impurity Drude 
conductivity in the framework of the method of
quantum kinetic equation.

In the 3D case the retarded Green's functions of scalar photons $V_{00}^R$ (the scalar Coulomb 
potential) and transverse vector photons $V_{11}^R$ in the Coulomb gauge 
$\nabla\cdot {\bf A}=0$ (${\bf A}$ is the vector potential) are given by the standard formulae \cite{R1,R2}
$$V_{00}^R(Q)={4\pi e^2\over q^2-4\pi e^2P_{00}^R(Q)},\ \ \ 
Q=({\bf q},\omega),                                            \eqno(1)$$
$$(V_{11}^R(Q))_{m,n}=V_{11}^R(Q)T_{m,n},  
  \ \ \ T_{m,n}=\delta_{m,n}-{q_mq_n\over q^2},  \eqno(2)$$
and $$V_{11}^R(Q)={4\pi e^2c^2\over \omega^2 -c^2q^2-
4\pi e^2c^2P_{11}^R(Q)};  \eqno(3)$$   
$m,n$ are the Cartesian indices and $c$ is the speed of light.
The scalar and the vector components of the vertex $a_{\mu}$ describing the electron-photon interaction are:
$$a_0=-1,  \ \ {\bf a}={1\over mc}\biggl({\bf p}+{{\bf q}\over 2}\biggr), 
                                                                \eqno(4)$$
where ${\bf p}$ is the electron momentum and $m$ is the electron mass.
\smallskip
The polarization operators for $ql>>1$ and $qv_F>>\omega$ are:
$$P_{00}^R(Q)=-\nu_0\bigl(1+i{\pi\omega\over 2qv_F}\bigr),
 \ \  P_{11}^R(Q)=-i{\pi \omega \nu_F v_F\over 4qc^2}, 
\ \ \nu_F={mp_F\over \pi^2},                                   \eqno(5)$$
where $v_F$ is the Fermi velocity, and $\nu_F$ is the two-spin 
electron density of states at the Fermi surface.
\smallskip
The bare retarded electron Green's function in an impure metal is
$$G_0^R(P)=(\epsilon -\xi_p+i/2\tau)^{-1}, \ \ \ P=({\bf p},\epsilon)   
\ \ \ \xi_p=(p^2-p_F^2)/2m.                                      \eqno(6)$$ 

In the Keldysh formalism \cite{K}, in addition to the retarded and advanced 
Green's 
functions, the more complicated electron $G^C$ and photon $V_{ii}^C$ Keldysh Green's
functions are introduced. Assuming that photons are in the thermodynamic 
equilibrium one can relate them as 
$$V_{ii}^C(Q)=(2N(\omega )+1)[V_{ii}^R(Q)-V_{ii}^A(Q)].            \eqno (7)$$
where $N(\omega)$ is the Bose distribution function. 

The electron system is considered to be driven out of equilibrium by an external  
electric field. Deriving the quantum kinetic equation
we make the conventional transformation from the coordinate to the 
momentum representation, the nonuniformity of the system being taken
into account by means of the corrections to the Poisson brackets.

In the lowest order in nonuniformity $G^C$ is given by the expression:
$$G^C(P)=S(P)[G^A(P)-G^R(P)]+\delta G^C(P),
\ \ \ \delta G^C(P)= {i\over 2}\lbrace S_0(\epsilon),
G^A(P)+G^R(P)\rbrace.                                            \eqno(8)$$
where the Poisson brackets in the presence of an electric field $\vec E$ are
$$\lbrace A,B\rbrace_E =e{\bf E}\biggl({\partial A\over \partial \epsilon}
{\partial B\over \partial{\bf p}}-
{\partial B\over \partial \epsilon}{\partial A\over \partial{\bf p}}\biggr), 
                                                                 \eqno(9)$$

The function $S(P)$ plays a role of the electron distribution function given
in the equilibrium by the formula $S=S_0=-\tanh(\epsilon/2T)$.
In the presence of the electric field $S$ is 
determined by the linearized quantum transport equation:
$$e({\bf v}\cdot{\bf E} ){\partial S_0\over \partial \epsilon}
=I_{e-imp}+I_{e-e},                                  \eqno(10)$$
where $I_{e-imp}$ and $I_{e-e}$ are the collision integrals which correspond
to the electron-impurity and the electron-electron scattering. The 
collision integrals are expressed in terms of the corresponding self-energies 
by virtue of the equations:
$$I(S)=I^0(S)+\delta I(S),\ \ \ I^0=-i[\Sigma^C-S(\Sigma^A-\Sigma^R)],$$
$$\delta I=-i[\delta \Sigma^C-S_0(\delta \Sigma^A-\delta \Sigma^R)]+
{1\over 2}\lbrace \Sigma^A+\Sigma^R,S_0\rbrace ,                 \eqno(11)$$
where $\delta \Sigma$ is the correction in the Poisson brackets form.
In our case $\delta \Sigma$ is obtained by taking into account the 
correction $\delta G^C$ in the expressions for $\Sigma$.
\smallskip
The collision integral $I_{e-imp}$ can be chosen in its simplest form:
$$I_{e-imp}={2\over \pi \nu \tau}\int {d{\bf k} \over (2\pi)^3}[S({\bf k},
\epsilon)-S({\bf p},\epsilon)]{\rm Im}G^A_0({\bf k},\epsilon)=
{S_0(\epsilon)-S(\epsilon)\over \tau}.                             \eqno(12)$$

Constructing the electron-electron collision integral we need
the retarded  electron self-energy 
$$\Sigma_{e-e}^R(P)=-\int {dQ\over (2\pi)^4}\biggl[ 
a_{\mu}V_{\mu\nu}^A(Q)a_{\nu}{\rm Im}[G^A(P+Q)]S(P+Q)$$
$$+a_{\mu}{\rm Im}[V_{\mu\nu}^R(Q)]a_{\nu} G^A(P+Q)(2N(\omega)+1)\biggr].      \eqno(13)$$
Assuming that the electron-impurity scattering is a dominant momentum relaxation 
process, we solve the transport equation (10) by iterations: $S=S_0+
\phi_0+\phi_1$, where $\phi_0$ is the first correction to the equilibrium 
distribution function $S_0$ which depends on the electron-impurity 
scattering but not on the electron-electron interactions
$$\phi_0({\bf p},\epsilon)=-\tau ({\bf v}\cdot {\bf E})
{\partial S_0(\epsilon)\over \partial \epsilon}.                 \eqno(14)$$

The next order correction $\phi_1$ includes the effects of the 
electron-electron interactions 
$$\phi_1=\tau[\delta I_{e-e}(S_0)]
={\tau\over 2}\lbrace\Sigma_{e-e}^A+\Sigma_{e-e}^R, S_0\rbrace$$
$$={\partial S_0(\epsilon)\over \partial \epsilon}\int{dQ\over (2\pi)^4}
\tau^2 e\biggl({\bf v}+{{\bf q}\over m}\biggr)\cdot{\bf E}$$
$${\rm Re}[a_{\mu}V_{\mu\nu}(Q)a_{\nu}]S_0(\epsilon+\omega){\rm Im}
\biggl(G_0^A(P+Q)\biggr)^2                                      \eqno(15)$$
In (15) we took into account only the nonlocal 
part of the collision integral from (11) which is expressed in terms of the
Poisson brackets. The reason is that only this part depends on the real
part of the exchange potential ${\rm Re}V_{ii}$ describing the interaction mediated by virtual photons.
In what follows we show that the processes involving virtual photons have a dominant effect on the conductivity renormalization.

The electric current is given by the equation
$${\bf J}_e=e\int{dP\over (2\pi)^4}{\bf v}{\rm Im}G^C(P),        \eqno(16)$$

We now treat  corrections to the Drude current due to the electron-electron 
interactions as the corrections to the distribution function and
the electron density of states
$$\Delta J_e=\delta\sigma {\bf E}=2e\int {dP\over (2\pi)^4}{\bf v}\biggl[
\phi_0{\rm Im}[(G_0^A)^2\Sigma_{e-e}^A(S_0)]+\phi_1{\rm Im}G_0^A$$
$$+S_0{\rm Im}\biggl((G_0^A)^2\Sigma_{e-e}^A(\phi_0)+
\phi_1{\rm Im}G_0^A\biggr)\biggr].                               \eqno (17)$$
The contributions of the first two terms in (17) cancel out. The contributions
of the third  and the fourth terms give
$$\Delta\sigma=2e^2\tau\int {dP\over (2\pi)^4}\int{dQ\over (2\pi)^4}
{\bf v}\cdot{\bf n}\biggl({\bf v}+{{\bf q}\over m}\biggr)\cdot
{\bf n}{\rm Re}[a_{\mu}V_{\mu\nu}(Q)a_{\nu}]$$
$$\biggl[S_0(\epsilon){\partial S_0(\epsilon+\omega)\over \partial \epsilon}
{\rm Im}(G_0^A(P))^2{\rm Im}G_0^A(P+Q)+$$
$$+S_0(\epsilon+\omega){\partial S_0(\epsilon)\over \partial \epsilon}
{\rm Im}G_0^A(P)
{\rm Im}\biggl(G_0^A(P+Q)\biggr)^2\biggr].                        \eqno(18)$$
Changing variables ${\bf p}\to {\bf p}+{\bf q}$ and 
$\epsilon\to\epsilon+\omega$
and then ${\bf q}\to-{\bf q}$ and $\omega\to- \omega$, one can see that the 
two terms in square brackets in (18) are identical.

\section{Gauge interactions in 3D}
Typically, in the case of gauge interactions relatively small momenta
transfers $q<<p_F$ are important. On the other hand, it can be readily seen that
the main contribution to the renormalization corrections
comes from photon momenta $q>>1/l$ and frequencies $\omega <v_Fq$. In this regime we first integrate in (18) the product  
of the electron Green's functions over $\xi_p$ and obtain
$$\int d\xi_p {\rm Im}G_0^A(P) {\rm Im}(G^A_0(P+Q))^2 =-{\pi^2\over (qv)^2}
\delta'\biggl(x-{\omega\over qv_F +{q\over 2p}}\biggr), \ \ 
x={{\bf p}\cdot{\bf q}\over pq}.                                    \eqno(19)$$
Next we perform the angular integration
$$\int {d\Omega_p\over 4\pi}\int {d\Omega_q\over 4\pi}{\rm Re}[a_{\mu}V_{\mu\nu}(Q)a_{\nu}]
\delta'\biggl(x-{\omega\over qv_F}+{q\over 2p}\biggr)
{\bf v}\cdot{\bf n}\biggl({\bf v}+{{\bf q}\over m}\biggr)\cdot
{\bf n}
=$$
$$={v_F^2\over 3}\biggl[ (-{q\over p_F}+{\omega\over qv_F})
\biggl({v_F\over c}\biggr)^2{\rm Re}V_{11}(Q)
+ (-{q\over 2p_F}){\rm Re}V_{00}(Q)  \biggr]
                   \eqno(20)$$
At this point one can notice the difference between the vector and the scalar couplings.
In the latter case the angular integral does not contain a $\omega$-term, 
and the absence of such a term dramatically reduces
the overall renormalization effect from the Coulomb interaction 
$V_{00}(Q)$ due to the parity reason (see later Eq.(22)).

Now we perform the integration over the photon momentum $q$ assuming that the $\omega$ -dependent
pole of the function $V_{11}(Q)$ is located at
$q> {\it max}(1/l, {\omega\over v_F})$ and end up with  
$$\int{dq q^2\over 2\pi^2}{\pi^2\over (qv_F)^2}{\omega\over qv_F} 
\biggl({v_F\over c}\biggr)^2{\rm Re}V_{11}(Q)=
-\biggl({v_F\over c}\biggr)^2{2\pi e^2\omega\over v_F^3}
{\pi\over 3^{3/2}(b|\omega|)^{2/3}}, \ \  b={\pi^2 e^2\nu_Fv_F\over c^2}.  \eqno(21)$$
The above assumption resulting in (21) requires $\omega$ to lie in the interval between $ T_1={1\over \tau(T_3\tau)^2}$ and 
$T_3={v_F^2\kappa\over c}$ (where $\kappa^2 =4\pi e^2\nu_F$), which is fairly broad in a clean metal.
For instance, for $E_F\sim 10 eV$, $v_F/c\sim 10^{-2}$, and $E_F\tau\sim 10^4$ we have 
$T_1\sim 10^{-3}K$, and $T_3\sim 10^3K$, so the condition $T_1<\omega <T_3$ 
is easy to satisfy in a wide range of $\omega$.

If, on the contrary, neither of the above conditions is met then the $q$ integral is determined by its lower limit,
which can be roughly estimated as either $q\sim 1/l$ or $q\sim {\omega/ v_F}$. Instead of (21) we then obtain
in the r.h.s. $\propto {\it min}(l^2, ({v_F\over \omega})^2)\biggl({v_F\over c}\biggr)^2{\pi e^2\omega\over v_F^3}$.

Now we are ready to carry out the frequency integrations where we have to distinguish between three different regimes
$\omega < T_1, T_1 <\omega <T_3,$ and $\omega <T_3$ (the intermediate regime exists only in the case of a pure metal $T_3 >>T_1$
which is equivalent to $E_F\tau >>({c^2\over e^2v_F})^{1/2}\sim 10^{2}$):
$$
{\Delta_{e-v\gamma}\sigma\over \sigma_0}=-{4e^2\over {\pi c^2 v_F}}[{\pi\over {3^{3/2}b^{2/3}}}\int_{T_1}^{T_3} d\omega \omega^{1/3}
f(\omega /T) +
{l^2\over 2}\int_{0}^{T_1}d\omega \omega f(\omega /T)+
{v^2_F\over 2}\int_{T_3}^{E_F}d\omega {f(\omega /T)\over \omega}]        \eqno(22)$$
where
$$f(\omega/T)={1\over 2}\int d\epsilon S_0(\epsilon+\omega)
{\partial S_0(\epsilon)\over \partial \epsilon}=
-{1\over 2}\int d\epsilon S_0(\epsilon)
{\partial S_0(\epsilon+\omega)\over \partial \epsilon}=
{\partial\over \partial \omega}[\omega\coth(\omega/2T)].      \eqno(23)$$
Note that $f(\omega)$ is an odd function, thus the finite result (22)
stems only from the term in Eq.(20) linear in $\omega$ which, in turn, originates from the vector
coupling to transverse gauge bosons.

For temperatures in the interval  $T_1 <T<T_3$ the leading $T$-dependent correction results from the first term inside the brackets
in Eq.(22) after a subtraction of the $T=0$ counterpart and  an extension of the $\omega$-integration from zero
to infinity.
Thus we finally obtain
$${\Delta_{e-v\gamma}\sigma\over \sigma_0}=
-{e^2v_F\over {\pi c^2}}\ln {c^2\over {e^2v_F}}
+
{2\pi^{1/3}\over 3^{5/2}}
\Gamma\biggl({4\over 3}\biggr)\zeta\biggl({4\over 3}\biggr)
\biggl({\kappa\over mc}\biggr)^{2/3}\biggl({T\over E_F}\biggr)^{4/3},\ \ \  \sigma_0=e^2{v_F^2\tau\over 3}\nu_F.
                                                                  \eqno(24)$$  
Notice that the overall correction to the conductivity remains negative whereas its variation with temperature is always
positive (the correction monotonically decreases in magnitude as $T$ increases).
Notably, both the constant and the $\sim T^{4/3}$ terms in Eq.(24) demonstrate
non-analytic dependences on the dimensionless 3D vector
coupling strength $(\kappa/mc)$.

At $T<<T_1$ the $T$-dependent contribution to ${\Delta_{e-v\gamma}\sigma\over \sigma_0}$ results from the second term in Eq.(22). It can be estimated by order of magnitude as 
$\sim {e^2l^2T^2\over v_Fc^2}$ whereas at high $T>> T_3$ it is due to 
the third term in Eq.(22) which yields $\sim {e^2v_F\over c^2}\ln {T\over E_F}$. 

In a pure metal these latter regimes are essentially irrelevant, 
although they do develop and
become more and more important as the amount 
of disorder increases. 

At last,
 in the dirty
limit $E_F\tau <10^2$ the first term in Eq.(22) 
and, correspondently, the temperature correction $\sim T^{4/3}$ disappear.
In this case the $T$-dependent contribution varies as 
$\sim {e^2l^2T^2\over v_Fc^2}$ at $T<<1/\tau$ and as 
$\sim {e^2v_F\over c^2}\ln {T\over E_F}$ at $1/\tau <<T<<E_F$.

The positive $\sim T^{4/3}$ term from Eq.(24) has to be compared with 
the other known conductivity corrections.
First, we compare it to the negative kinetic term 
${\Delta^{'}_{e-v\gamma}\sigma/
\sigma_0}\propto -({\kappa\over {mc}})^{10/3}{\tau T^{5/3}\over E^{2/3}_F}$, 
which represents the relativistic electromagnetic interaction correction to the transport scattering rate and
accounts for the exchange of real transverse photons \cite{R1}.
The kinetic contribution to $\Delta^{'}_{e-v\gamma}\sigma (T)$ vanishes at zero temperature and is proportional to $\sigma^2_0$
while the renormalization term (24) is only of the first order in  $\sigma_0$. However, the high power of the small parameter
$(\kappa/mc)$ present in the kinetic term guarantees that the temperature dependence of the measured conductivity 
is dominated by the renormalization term at all $T < \biggl(E^2_F\tau^3 (\kappa/mc)^8\biggr)^{-1}$.
The latter condition is easy to satisfy unless the system is in the extremely clean limit $E_F\tau >10^5-10^6$.

The next conductivity correction, which Eq.(24)
has to be compared with,
is the well-known Altshuler-Aronov term  originating from Coulomb exchange
processes in the diffusive regime $ql<1$ \cite{AA}. This quantum interference
correction, which is only relevant at $T<1/\tau$,
has the same negative sign as Eq.(24) while it increases with temperature:
${\Delta_{AA}\sigma/
\sigma_0}=-c_1(E_F\tau)^{-2}+ c_2(E_F\tau)^{-3/2}(T/E_F)^{1/2}$,
where $c_{1,2}$ are positive constants. 

In fact, at all $T<E_F (E_F\tau)^{-9/5}(\kappa/mc)^{-4/5}\sim 10^{-1}K$ the temperature
dependent part of $\Delta_{AA}\sigma$ exceeds the new 
$\sim T^{4/3}$ term from the interference correction
(24) resulting from processes with momenta transfers $ql>1$.
Nevertheless, at $T=0$ the conductivity
correction is controlled by the new term (24) rather than by the 
Altshuler-Aronov term unless $E_F\tau <10^2$.
 
Since both $\Delta_{AA}\sigma$ and $\Delta_{e-v\gamma}\sigma$ given by Eq.(24)
increase  as a function of temperature, one could assume that at all feasible 
temperatures the overall conductivity of a system
of 3D fermions weakly coupled to gauge bosons increases as well! 

However, in a real metal the (negative) kinetic electron-phonon interaction correction 
${\Delta_{e-ph}\sigma/
\sigma_0}\propto
 -{\tau T^2\over E_F}$ \cite{R1} dominates over the Altshuler-Aronov term
 at all $T> E_F (E_F\tau)^{-5/3}\sim 10^{-2}K$ and over the new term (24) at all   
$T> E_F ({\kappa\over mc})(E_F\tau)^{-3/2}\sim 10^{-4}K$
for the typical parameter values. 

For the sake of completeness we note
that in a metal 
$\Delta\sigma$ receives another contribution coming from interference between the electron-impurity and the electron-phonon
scattering, which was  estimated in \cite{RS1} as ${\Delta_{e-ph-imp}\sigma/
\sigma_0}\propto -{T^2\over E_Fp_F u_l}$, where $u_l$ is the longitudinal sound velocity. Although formally
this term receives contributions of both signs, its actual value is
negative, given the fact that the longitudinal sound
has higher velocity than the transverse one $(u_l >u_t)$.

It is also worthwhile mentioning
that in a ferromagnetic metal with a high magnetic permuability the electron interaction with transverse vector photons
is strongly amplified  \cite{RT}, which makes the above renormalization effects essentially more pronounced.

As the last remark, we presume that our results can be also used in the analysis of transport properties
of a hot relativistic quark-gluon plasma \cite{BPM}.

\section{Gauge interactions in 2D}
Next we consider the $ql>1$ conductivity renormalization in the case of 2D fermions coupled to scalar (longitudinal)
and vector (transverse) gauge bosons
whose  
propagators are given by the formulae (1-5)   
where  instead of $e^2$ we now use the notation $g^2$ for the coupling constant with the dimension of energy,
and the two-spin density of states is $\nu^{2D}_F={m\over \pi}$. The velocity of gauge bosons $c$ is another parameter
which may well be  comparable with $v_F$, especially if the above gauge theory serves as some sort of
an effective description of an underlying non-relativistic microscopic Hamiltonian
and the role of the gauge boson
is played by one of the collective modes. To avoid a possible confusion we stress that the dynamics of gauge bosons is purely
two-dimensional, so that the 2D scalar
 interaction is completely screened in the quasi-static limit $\omega <<v_Fq$: $V_{00}(Q)\approx {4\pi g^2\over {q^2+\kappa^2}}$,
where $\kappa^2 =4\pi g^2\nu^{2D}_F$. The 2D 
conductivity which we are going to compute, is the response to an ordinary "in-plane" electric field coupled
in the usual way to our 2D fermions carrying the electric charge $e$ in addition to the 
2D gauge coupling $g$.  

First we compute the renormalization correction resulting from the transverse vector coupling.
The angular integral analogous to Eq.(20) now reads as
$$\int {d\Omega_p\over 2\pi}\int {d\Omega_q\over 2\pi}{\rm Re}[a_{\mu}V_{\mu\nu}(Q)a_{\nu}]
\delta'\biggl(x-{\omega\over qv_F}+{q\over 2p}\biggr)
{\bf v}\cdot{\bf n}\biggl({\bf v}+{{\bf q}\over m}\biggr)\cdot
{\bf n}
=$$
$$={v_F^2\over 2\pi}\biggl[ (-{3q\over 2p_F}+{\omega\over qv_F})
\biggl({v_F\over c}\biggr)^2{\rm Re}V_{11}(Q)
-({q\over 2p_F}+{\omega\over qv_F}){\rm Re}V_{00}(Q)  \biggr]  
                             \eqno(25)$$

Instead of Eq.(22) we now obtain
$$
{\Delta_{vec}\sigma^{2D}\over \sigma^{2D}_0}=-{4 g^2\over {3\pi c^2 v_F}}
[{\pi\over {2b_{2D}}}\int_{T_1}^{T_3} d\omega 
f(\omega /T) +
l^3\int_{0}^{T_1}d\omega \omega f(\omega /T)+
{v^3}\int_{T_3}^{\infty}d\omega {f(\omega /T)\over \omega^2}]          \eqno(26)$$
where $b_{2D} ={2\pi g^2v_F\nu^{2D}_F\over c^2}$ and $\sigma^{2D}_0={1\over 2}e^2v^2_F\tau\nu^{2D}_F$.

In the range of temperatures $T_1 <T<T_3$ which exists if coupling is strong enough (${g^2\over mc^2}>(E_F\tau)^{-2}$)
the vector contribution is given by
$${\Delta_{vec}\sigma^{2D}\over \sigma^{2D}_0}=
-{2^{1/2}\over {3\pi}}\bigl({\kappa\over mc}\bigr)(1+{\pi\over 2})
+
{T\over 6E_F}          \eqno(27)$$ 
Note, that the dependence on the coupling constant $g$ is again non-analytic,
in particular, the coefficient in front of the $\sim T$ term  does not contain
$g$ at all. 
Eq.(27) has to be contrasted with the 2D counterpart of the (negative)
kinetic term ${\Delta{'}_{vec}\sigma^{2D}/ \sigma^{2D}_0}\propto -{\tau T^{4/3}\over E^{1/3}_F}\biggl({\kappa\over mc}\biggr)^{8/3}$,
which corresponds to the correction to the transport scattering rate.

At $T<<T_1$ the r.h.s. of (26) varies with temperature as $\sim {g^2l^3T^2\over vc^2}$ while at $T>>T_3$ it decays as
$\sim  -{g^2v^2_F\over c^2T}$.

It is also worthwhile mentioning that in the case of weak coupling ( ${g^2\over mc^2}<(E_F\tau)^{-2}$)
the constant term is $\sim -{g^2v^2_F\tau\over c^2}$
and there are only two latter regimes for the $T$-dependent part which match together
at $T\sim 1/\tau$.

Remarkably, in the 2D case the 
scalar potential ${\rm Re}V_{00}(Q)$ also leads to a significant contribution 
associated with the $\omega$-odd term resulting from the 2D angular integral (25)
$$
{\Delta_{sc}\sigma^{2D}\over \sigma^{2D}_0}=-{1\over {\pi^2mv^3_F}}
[\int_{0}^{1/\tau} d\omega \omega 
f(\omega /T)(l-{1\over \kappa}\tan^{-1}\kappa l) +
\int_{1/\tau}^{\infty}d\omega {f(\omega /T)}(v_F-{\omega\over \kappa}\tan^{-1}{\kappa v_F\over \omega})]          \eqno(28)$$
At $g^2 > 1/\tau (E_F\tau)$ there exists a range ot temperatures $1/\tau <T<T_3\sim (E_F g^2)^{1/2}$ where 
$${\Delta_{sc}\sigma^{2D}\over \sigma^{2D}_0}=
-{\it const}
+ {T\over \pi E_F}
                                                                  \eqno(29)$$
and the constant term behaves as $\sim (g^2/E_F)^{1/2}$. The term linear in $T$ comes without any smallness just as in the
vector case. Notice that as compared to the vector case
 which we considered above
the scalar vertices are free of the factor $(v_F/c)^2$. 

At $T<<1/\tau$ the temperature dependent part is again quadratic 
$\sim {\tau T^2\over E_F}$ and independent of coupling whereas at $T>>T_3$ it now behaves as $\sim -g^2/T$.

In the weak coupling case  $g^2 < 1/\tau (E_F\tau)$
the constant term in Eq.(28) is $\sim -g^2\tau$ and there are only 
two different regimes exhibited by the $T$-dependent part of the r.h.s. of Eq.(28):
$\sim g^2\tau^3T^2$ at $T<1/\tau$ and $\sim -g^2/T$ in the opposite case.

The expressions (26) and (28) have to be compared with the Altshuler-Aronov-type
2D interference corrections resulting from the diffusive regime $ql<1$, which appear to be independent of the coupling strength at $T<1/\tau$ and diverge 
logarithmically as $T$ tends to zero.
It turns out that the 2D vector gauge interaction produces a negative
term ${\Delta_{AA}\sigma^{2D}/ \sigma^{2D}_0}\sim -(E_F\tau)^{-1}|\ln(E_F\tau)|^{1-n}|\ln(T\tau)|^{1+n}$,
where the exponent $n$ is either 0 or 1 depending on whether $T<1/\tau (E_F\tau)^{-2}$
or $1/\tau (E_F\tau)^{-2}<T<1/\tau$ (see also \cite{MW}). 

In the case of the 2D scalar gauge interaction
the diffusive regime yields a subdominant term of order $-(E_F\tau)^{-1}|\ln(T\tau)|$ at all $T<1/\tau$ (a similar term arises due to the effects of weak localization,
where the phase relaxation time due to 2D gauge interactions is given by 
$\tau^{-1}_{\phi}\sim T^{1/3}$ \cite{AW}).
The low temperature divergence of the first order $ql<1$ interference and localization
corrections requires a tedious
account of higher order terms, which have not been done yet. 

\section{Discussion}
In the modern Condensed Matter Theory the effective description in terms of 2D gauge fields arises in a number of contexts.
The well known examples are the gauge theory of the doped Mott insulators, which is believed to be relevant
for the problem of the high $T_c$ superconductivity \cite{gauge}, and the gauge theory of half filled Landau level \cite{HLR}.  

Unfortunately, neither of these problems
features an example of the weak gauge coupling regime in cases of physical interest.
Therefore our results based on the perturbative solution of the quantum kinetic equation under the assumption
of the dominant impurity scattering can not be used directly in these contexts.

Moreover, the present gauge theory of the normal state of high $T_c$ cuprates
\cite{gauge} involves two kinds of excitations (spinons and holons) 
coupled to the gauge field, the physical
conductivity being
governed by that of spinless charged bosons (holons).
 Given all these complications, we would like to warn against
any attempt of a direct use of the formulae (27) and (29) in the context of linear
resistivity of high $T_c$ materials, which was explained in \cite{gauge}
by the $\sim T$ behavior of the standard 
(kinetic) holon-gauge boson scattering rate.

Nevertheless, our analysis implies, for instance, that the gauge interaction of spinons in doped Mott insulators \cite{gauge}
strongly affects the classical impurity conductivity even at low temperatures. The $T=0$ renormalization factor may well be 
of order unity, the fact to be kept in mind at an attempt to make any quantitative predictions. 

In the case of half filled Landau level there is another reason why one can not straightforwardly apply the above results
even in the artificial limit of small $\Phi$, the number of flux quanta attached to every electron 
(the physical case corresponds to $\Phi =2$, of course). It was pointed out elsewhere \cite{DVK} that despite the external magnetic 
field gets cancelled in average by the attached flux, 
 the dynamics of new fermionic quasiparticles (named composite fermions \cite{HLR}) remains diffusive
up to transferred momenta $q\sim 1/l_B=B^{1/2}$. Therefore in the composite fermion theory
there is no room for the ballistic regime $"ql>1"$ which we discuss in the present
paper.

To add to this point, we mention that if it were not the case, then the low-temperature conductivity, which is governed by the
divergent (negative) diffusion correction, would manifest the
$(\ln T\tau)^2$ behavior in the 
interval $(E^2_F\tau^3)^{-1}<T<1/\tau$  \cite{MW}. 
Such a prediction
 would certainly contradict the
experiment \cite{log}, which  demonstrates the $\ln T$ behavior of the conductivity $\sigma_{xx}(T)$ 
at filling factors $\nu =1/2$ and 3/2 at temperatures from $0.5 K$ down to $15 mK$,
while the estimates based on the parameters of samples used in \cite{log} yield
$1/\tau \approx 0.5K$ and $(E^2_F\tau^3)^{-1}\approx 10mK$. 

On the other hand,  assuming that the diffusion of composite fermions extends up
to distances of order of the magnetic length $l_B\sim B^{-1/2}$,
one finds the leading $\ln T$ behavior of $\sigma_{xx}(T)$ in the whole range
of temperatures $(E^2_F\tau^3)^{-1}<T<1/\tau$ \cite{DVK}.
It is worthwhile to note that in the original electron picture the existence of diffusion at the magnetic length 
scale
can be readily seen from the fact that the diffusive behavior results from electron hoppings between adjacent Landau orbitals, which are
$l_B$ distance apart (this fact becomes much more obscure after a mapping of electrons onto composite fermions though). 

As another implicit evidence supporting our arguments we mention a similar effect of the vector gauge interactions
on the renormalization of thermopower, which was previously discussed in the contexts of the 3D electron-phonon \cite{RS}
and the 3D electron-electron interaction \cite{R3} problems. In the case of 
 thermopower the leading $ql>1$ correction to the thermoelectric coefficient $\eta$
is given by the expression similar to Eq.(18) but with an extra factor $\epsilon/T$ in the integrand. Therefore one only needs
the $\omega$-even part of (20) or (25), which is nonzero for both vector and scalar couplings.

Then in the (obviously 2D) case of  half filled Landau level the $ql>1$
correction to the Drude thermopower $(S_0\approx {\pi^2T\over 3eE_F})$
of electrons with the unscreened Coulomb potential
 would behave as $\Delta S \sim {T\Phi^2k_F\over e^3mE_F}\log {E_F/T}$ (in the screened case it becomes even stronger
$\Delta S_{xx} \sim {1\over e}({T\Phi^2\over E_F})^{2/3}$ in accordance with the general expectations \cite{HLR}).
Although one can not simply interpolate these perturbative corrections into the physical case of $\Phi =2$, 
they indicate a possible strong nonlinear $T$ dependence of the measured thermopower. 
However, the available experimental data for the diffusion thermopower $S(T)$
at even denominator fractions \cite{TEP} do not seem to support the existence of any substantial non-linear  terms. 
Moreover, they rather indicate that the corrections to $\sigma_0$ are relatively small
(see also \cite{DVK2}). 
This observation is consistent with the absence of the ballistic regime (and the related $ql>1$ renormalization effects)
in the composite fermion 2D gauge theory. 

Our analysis of the 2D scalar case also implies that similar renormalization effects due to the ordinary (non-gauge)
electron-electron
and electron-phonon interactions do occur,
for example, in doped semiconductor heterostructures. 

In contrast to the above examples of 2D gauge theories
the conventional Coulomb interaction $V_{00}(Q)={2\pi e^2\over {\epsilon_0(q+\kappa)}}$ (where $\kappa =2\pi e^2\nu_{2D}/\epsilon_0$
contains the dielectric constant $\epsilon_0$)
 may indeed feature a small parameter $\alpha =\kappa/2p_F$ at high enough sheet electron densities.

Repeating the calculations, which led to (28), we now obtain 
$$
{\Delta_{Coul}\sigma^{2D}\over \sigma^{2D}_0}=-{1\over {\pi mv^3_F}}
[\int_{0}^{1/\tau} d\omega \omega 
f(\omega /T)(l-{1\over \kappa}\ln (1+\kappa l)) +
\int_{1/\tau}^{\infty}d\omega {f(\omega /T)}(v_F-{\omega\over \kappa}\ln (1+{\kappa v_F\over \omega}))]          \eqno(30)$$

Again, at $\kappa l>>1$ there exists a range ot temperatures $1/\tau <T<T_3=\kappa v_F$ where 
$${\Delta_{Coul}\sigma^{2D}\over \sigma^{2D}_0}=
-{1\over 2\pi}\int^{\Omega/E_F}_{0}dx [1-{x\over 4}-{x\over 4\alpha}\ln\biggl({x+4\alpha\over {x(1+\alpha)}}\biggr)]
+  {T\over \pi E_F}
                                                                  \eqno(31)$$
The constant term in (31) yields zero temperature renormalization, which depends on both
the upper frequency cutoff $\Omega$ and the coupling strength $\alpha$.
We estimate the constant term at small $\alpha$ as $-{\kappa\over 2\pi p_F}\ln({\Omega\over \kappa v_F})$
while in the strong screening limit ($\alpha >>1$) it approaches the value $-{\Omega\over 2\pi E_F}(1-{\Omega\over 8E_F})$.

Thus, despite of an uncertainty of the actual value of $\Omega\sim E_F$ we find
that the renormalization effect might become quite substantial at low densities, which correspond
to large values of $\alpha$.
It is worthwhile mentioning that the situation  $\alpha\sim 1$ arises, for example, in low-density ( $n_e <10^{11}cm^{-2}$)
$GaAs$ heterostructures characterized by the dielectric constant
$\epsilon_0=13$ and the electron band mass $m=0.067m_0$.
In this system the renormalization correction (31) could remain greater
than the 2D Altshuler-Aronov term or the localization correction
(in the presence of Coulomb interactions the latter is governed by the 2D phase relaxation time $\tau^{-1}_{\phi}\sim T$ \cite{AAL}),
both given by the expression
${\Delta_{AA}\sigma^{2D}/ \sigma^{2D}_0}={\Delta_{wl}\sigma^{2D}/\sigma^{2D}_0}=
-{1\over \pi(E_F\tau)}|\ln(T\tau)|$, except at extremely low temperatures.

At $T<<1/\tau$ the temperature dependent part of (30) is again quadratic 
$\sim {\tau T^2\over E_F}$ and independent of coupling whereas at $T>>\kappa v_F$ it now behaves as
 $\sim -{\kappa\over p_F}\ln {\Omega\over T}$.

In the strongly disordered case  $\kappa l<1$ (which is, however, not quite important for the analysis of
typically clean $GaAs$ samples) there are only 
two latter regimes left over. 

Among other possible applications of the above results we mention 
the problem of transport properties in the vicinity of 
a quantum critical point corresponding to some charge transfer instability.
In this scenario, which was argued to be relevant for the problem of
high $T_c$ cuprates \cite{V},  charged
fermionic excitations are coupled in a scalar way to
an overdamped critical mode described by the
propagator $(i\omega/q^{\alpha}+q^{\beta})^{-1}$. Our results suggest that at
a crossover temperature, above which the critical fluctuations become effectively
three-dimensional, the conductivity corrections get strongly suppressed 
as compared to those in the low temperature (effectively two-dimensional) regime.    

\section{Summary}
To summarize,
in the present paper we show that the Drude conductivity of fermions coupled to three- or two-dimensional gauge fields
is strongly renormalized due to corrections to the nonequilibrium fermion density of states and the nonlocal part 
of the electron-electron collision integral. We also correct the earlier prediction \cite{SRW} of a similar renormalization effect
in the 3D electron-phonon problem. To this end, we demonstrate that it is a pecularity
of the relevant phase volume-type expression in the case of the 3D scalar
(density-density) coupling which makes this new correction negligibly small.
However, in the case of the vector (current-current) coupling such terms do appear in both 2D and 3D.
At strong enough coupling which, however, still remains in the perturbative regime the temperature dependent part of the correction
is found to behave as $T^{4/3}$ in the 3D and as $T$ in the 2D case (in the latter case the scalar coupling contributes as well, and the coefficient in front of the $\sim T$ term appears to be independent of the coupling strength).
 The non-analytic dependences of these corrections resulting from
large momenta transfers ($ql>1$) on the coupling strength and/or temperature
allow one to classify them as a non-Fermi liquid renormalization.

D.V.K. acknowledges the hospitality extended to him 
at the Aspen Center for Physics where this work was completed.
M.R. acknowledges the hospitality at the Institute for Theoretical Physics at
Santa Barbara where part of this work was done.
The research of M.R. was supported in part by NSF Grant No. PHY94-07194
and US DOE Contract No. DE-FG02-ER45347.

\eject

\end{document}